\documentclass{article}

\usepackage[nonatbib, preprint]{neurips_2026}

\usepackage[utf8]{inputenc} 
\usepackage[T1]{fontenc}    
\usepackage{url}            
\usepackage{amsmath,amssymb,amsthm}
\usepackage{booktabs}       
\usepackage{amsfonts}       
\usepackage{nicefrac}       
\usepackage{hyperref}       
\hypersetup{colorlinks=true, citecolor=magenta, linkcolor=blue, urlcolor=black}
\usepackage{graphicx}
\usepackage{microtype}      
\usepackage[numbers]{natbib}
\usepackage[most]{tcolorbox}
\usepackage{multirow}
\usepackage{algorithm}
\usepackage{algorithmic}
\usepackage{enumitem}
\usepackage{fvextra}  
\usepackage{xspace} 
\usepackage{subcaption}

\DeclareMathOperator*{\argmin}{arg\,min}
\newcommand{\AgenticTuner}{\textsc{AgenticTuner}\,}

\usepackage{xcolor}

\newcommand{\Pfile}{\mathcal{P}}
\newcommand{\Ccode}{\mathcal{C}}

\newcommand{\Guide}{\mathcal{G}}
\newcommand{\Dhist}[1][t]{\mathcal{D}_{#1}}
\newcommand{\Notebook}{\mathcal{N}_t^{(i)}}
\title{Agentic Systems for Calibrating Cosmological Simulations}

\author{%
  Chaipat Tirapongprasert \\
  Columbia University\\
  \texttt{t.chaipat@columbia.edu} \\
  \And
Matthew Ho\\
Institute for Advanced Study
}

\newcommand{\pbbold}[1]{{\sffamily\bfseries #1}}
\newcounter{promptboxctr}
\newtcolorbox{promptbox}[1]{
  enhanced, breakable, colback=gray!4, colframe=gray!55!black,
  boxrule=0.4pt, arc=1pt, left=6pt, right=6pt, top=4pt, bottom=4pt,
  fonttitle=\bfseries\small, title={Box~\thepromptboxctr: #1},
  fontupper=\ttfamily\small
}

\begin{document}

\maketitle

\begin{abstract}
Calibrating a cosmological simulation is an expensive, high-dimensional parameter optimization task. Previous works employ Bayesian optimization for automation, which treats the simulator as a black box and precludes the incorporation of human domain knowledge or workflow into the selection policy. We propose \AgenticTuner, an agentic calibration pipeline with a present-day large language model serving as the optimizer who proposes the next parameter trial. We show that these agents can achieve competitive calibration performance at a fraction of the CPU cost when they are given informal, human-readable priors (such as codebase, documentation, and heuristic guides that seasoned scientists may possess). We hope this work serves as an early step toward AI scientists that can carry out complex, long-horizon scientific tasks. A starter code is available on GitHub.\footnote{\url{https://github.com/chaipattira/SimulationCalibrator}}
\end{abstract}

\section{Introduction}

Cosmological hydrodynamic simulations necessitate a set of physical parameters that regulate processes like stellar and active galactic nuclei (AGN) feedback. These parameters cannot be determined from first principles and must be adjusted for the simulation to reproduce known observables. To produce simulations and emulators that generalize across plausible physics, large simulation suites, such as CAMELS \citep{villaescusa2021camels} and DREAMS \citep{rose2025introducing}, vary parameters over wide ranges. The problem is that many of these parameters, particularly those governing astrophysical feedback, lack strong physical priors, making it unclear how to center the sampled ranges around values that are consistent with actual observations.

To produce a well-calibrated simulation, practitioners can either manually iterate through a trial-and-error tuning procedure based on prior knowledge and experience or automate the search with Bayesian optimization~\citep[BO;][]{rasmussen2006gaussian,akiba2019optuna, rogers2019bayesian, leclercq2018bayesian}.
Despite its effectiveness, BO cannot easily absorb the vast corpus of domain knowledge already present in natural language such as scientific papers or simulation code.
Customized kernels and surrogates ~\citep{feurer2015initializing, hvarfner2022pibo} can be used to encode structural assumptions from documentation and previous experimental logs, but these methods are challenging to formalize and implement.
Recent research ~\citep{zhang2023llmhpo,liu2024llambo,madiraju2025optimindtune} proposed replacing the acquisition function with large language models (LLMs), prompting them to propose the next query point conditioned on a textual description of the problem and search space. However, such direct prompting is inadequate for rich, long-horizon tasks, such as ours, which require a large amount of relevant context that cannot be accommodated in a single context window.

In this paper, we present \AgenticTuner, an agentic simulation calibration pipeline in which an LLM agent can reason about the search landscape, form hypotheses, and retrieve the relevant domain knowledge priors. Like a human scientist, it can \textbf{(a)} consult documentation, source code, or a running record of previous experiments and \textbf{(b)} query low-resolution proxy simulation to narrow down a portion of the parameter space before committing to an expensive run.
Our findings show that \AgenticTuner can match the calibration performance of a Bayesian optimization baseline at a fraction of the simulation cost, demonstrating the practical value of encoding domain priors into the selection policy. The agents can also adjust their strategies based on budget and evaluation history.

\section{Problem and Experimental Setup}

\paragraph{Task.}
We approach the calibration of an expensive, intractable-likelihood simulator~\citep{kugel2023flamingo, hacc2026emulator,rogers2019bayesian} as an optimization problem.
For a chosen parameter vector $\theta_{\rm true} \in \Theta \subset \mathbb{R}^{d}$, we run a forward simulator $\mathcal{M}$ to generate a target observable $x^{\rm target} = \mathcal{M}(\theta_{\rm true})$. The goal of calibration is to search for a point estimate
\begin{equation}
  \hat\theta \;=\; \argmin_{\theta \,\in\, \Theta } \; d\big(\mathcal{M}(\theta),\, x^{\rm target}\big)
  \quad \text{s.t.} \quad \textstyle\sum_t c(\theta_t) \,\le\, B\,,
  \label{eq:calibration-objective}
\end{equation}
where $d(\cdot,\cdot)$ is a discrepancy metric defined via the statistical goodness-of-fit of the observable and $c(\theta)$ is the computational cost of running one forward pass of the simulator. As a constraint, we specify that the total cost of running these trials (e.g., CPU hours, wall-clock time) cannot exceed a predetermined budget $B$. For cosmological simulators, $\mathcal{M}$ is typically a very complex non-linear mapping, making Eq. \ref{eq:calibration-objective} a non-convex optimization objective.

\begin{figure}
    \centering
    \includegraphics[width=1\linewidth]{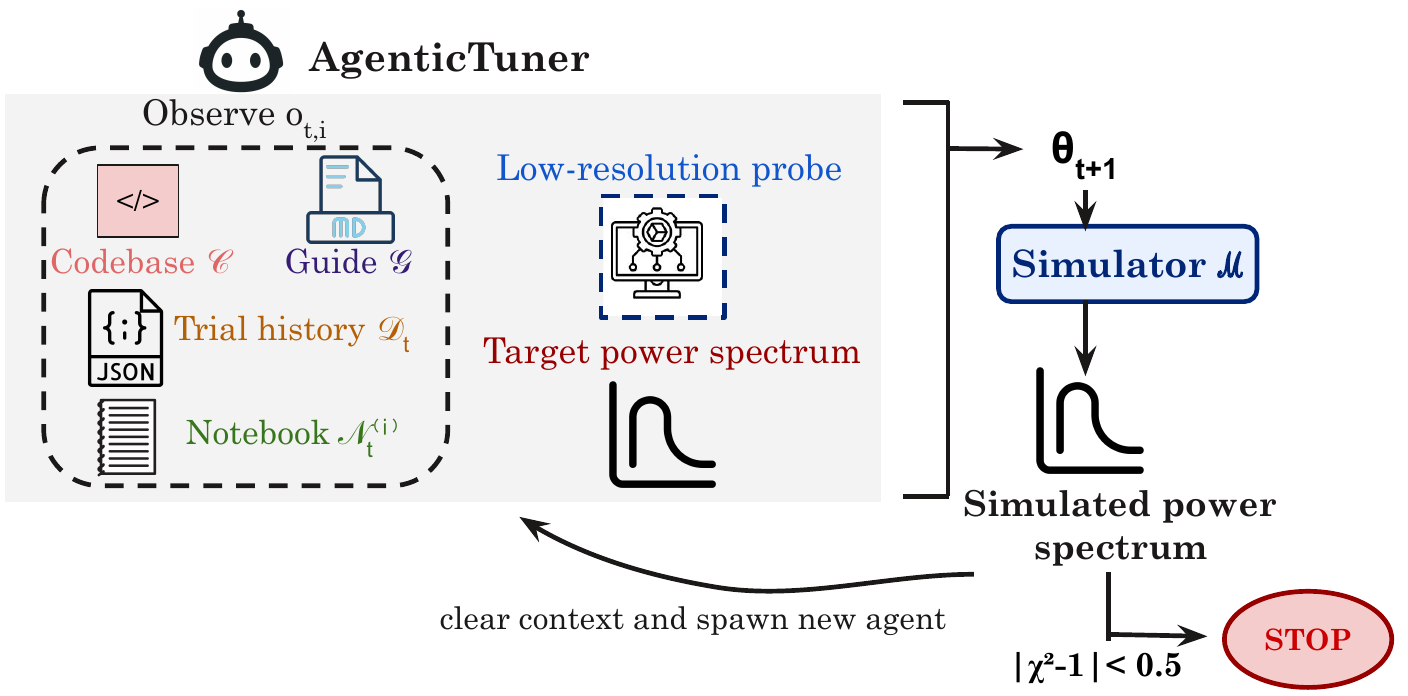}
\caption{Overview of \textsc{AgenticTuner}. At each step $t$, the agent can observe the workspace $o_{t,i}$ or run a low-resolution probe before submitting $\theta_{t+1}$ to the simulator $\mathcal{M}$. If the trial converges, the loop terminates; otherwise, the agent's context is cleared, and the loop restarts with the updated $\mathcal{D}_t$.}
\label{fig:agentictuner}
\end{figure}

\paragraph{Baseline.}
A standard non-convex optimization technique is BO~\citep{rasmussen2006gaussian,akiba2019optuna}, which treats Eq.~\ref{eq:calibration-objective} as a sequential decision process. At each step $t$, BO fits a probabilistic surrogate model to the trial history $\mathcal{D}_t = \{(\theta_\tau, d_\tau, c_\tau)\}_{\tau \le t}$, and uses an acquisition function $\alpha(\theta; \mathcal{D}_t)$ to propose the next point, $\theta_{t+1} = \arg\max_{\theta \in \Theta} \; \alpha(\theta; \mathcal{D}_t)$. The most common surrogate is a Gaussian process~\citep[GP;][]{rasmussen2006gaussian}, with tree-based and density-based alternatives also in use (e.g., TPE~\citep{bergstra2011algorithms}, SMAC~\citep{lindauer2022smac3}). Regardless of surrogate choice, BO treats the optimization process as a black box with a non-informative prior, which prevents the integration of domain-specific knowledge or workflow into the selection policy. We choose GP as the baseline because it was the most sample-efficient method in preliminary experiments comparing multiple derivative-free optimizers for this class of problem. Appendix~\ref{app:sim-details} outlines our baseline implementation of GP-BO.

\paragraph{Configuration.}
We center our study on a sufficiently challenging numerical task of calibrating a cosmological simulation. We use MP-Gadget \citep{bird2022astridI, ni2022astridII}, a particle-based (N-body) cosmological simulation that evolves dark matter, gas, stars, and black holes via gravity and hydrodynamics, as the forward simulator $\mathcal M$.
For a chosen 12-parameter vector $\theta_{\rm true}$ (defined in Table~\ref{tab:params}), $\mathcal{M}$ generates the target observable $x^{\rm target}$ from Eq.~\ref{eq:calibration-objective} in the form of a matter power spectrum
$P^{\rm target}(k;\theta_{\rm true}) \in \mathbb{R}^{n_{\rm bins}}$: a statistic measuring matter clustering strength as a function of scale $k$, evaluated at $n_{\rm bins}$ Fourier bins. For a given simulation configuration, this vector is expected to have intrinsic noise distributed as an i.i.d. isotropic Gaussian. Appendix~\ref{app:sim-details} provides more details on simulation and job settings.

We instantiate the discrepancy $d(\cdot, \cdot)$ from Eq.~\ref{eq:calibration-objective} as the mode-weighted reduced chi-squared between simulated and target spectra,
\begin{equation}
    \tilde\chi^2 = \frac{\chi^2}{n_{\rm bins}},\qquad \text{where} \qquad \chi^2 = \sum_i \frac{\big(P_i^{\rm sim} - P_i^{\rm target}\big)^2}{\sigma_i^2},
\end{equation}
and $\sigma_i^2$ is the cosmic variance, \textit{i.e.} the statistical noise from measuring clustering in a finite volume (full expression and $k$-range in Appendix~\ref{app:sim-details}). A trial is considered converged when $\tilde\chi^2<1$, i.e., once the simulated spectrum matches the target to within the cosmic-variance noise floor. The CPU hours spent on trial $t$ determine its cost $c(\theta_t)$ from Eq.~\ref{eq:calibration-objective}.

\section{Methods}
\label{sec:agent}

\paragraph{Loop design.} \AgenticTuner is an agentic simulation calibration system that replaces an acquisition function with an LLM as the optimizer (see Figure~\ref{fig:agentictuner}).
The central loop of \AgenticTuner is orchestrated by the external harness, which is separate from and independent of the LLM. Each iteration $t$ begins with the instantiation of a stateless LLM instance within a sandboxed container that conceals the hidden ground truth $\theta_{\rm true}$. Because LLM performance has been shown to deteriorate with a monotonically increasing context window~\citep{dongre2025drift,xia2026diagnosing,wang2026horizon}, we opt to clear the context~\citep{anthropic2025context} after the LLM calls an expensive simulator for the proposed trial $\theta_{t+1}$.
When the simulation is finished, the harness mediates the evaluation process by calculating $\tilde\chi^2(\theta_{t+1})$ and either restarting the loop with a fresh agent or stopping the search if $\tilde\chi^2<1$ or a fixed CPU-hour budget is exhausted.

\begin{figure}
    \centering
    \includegraphics[width=\linewidth]{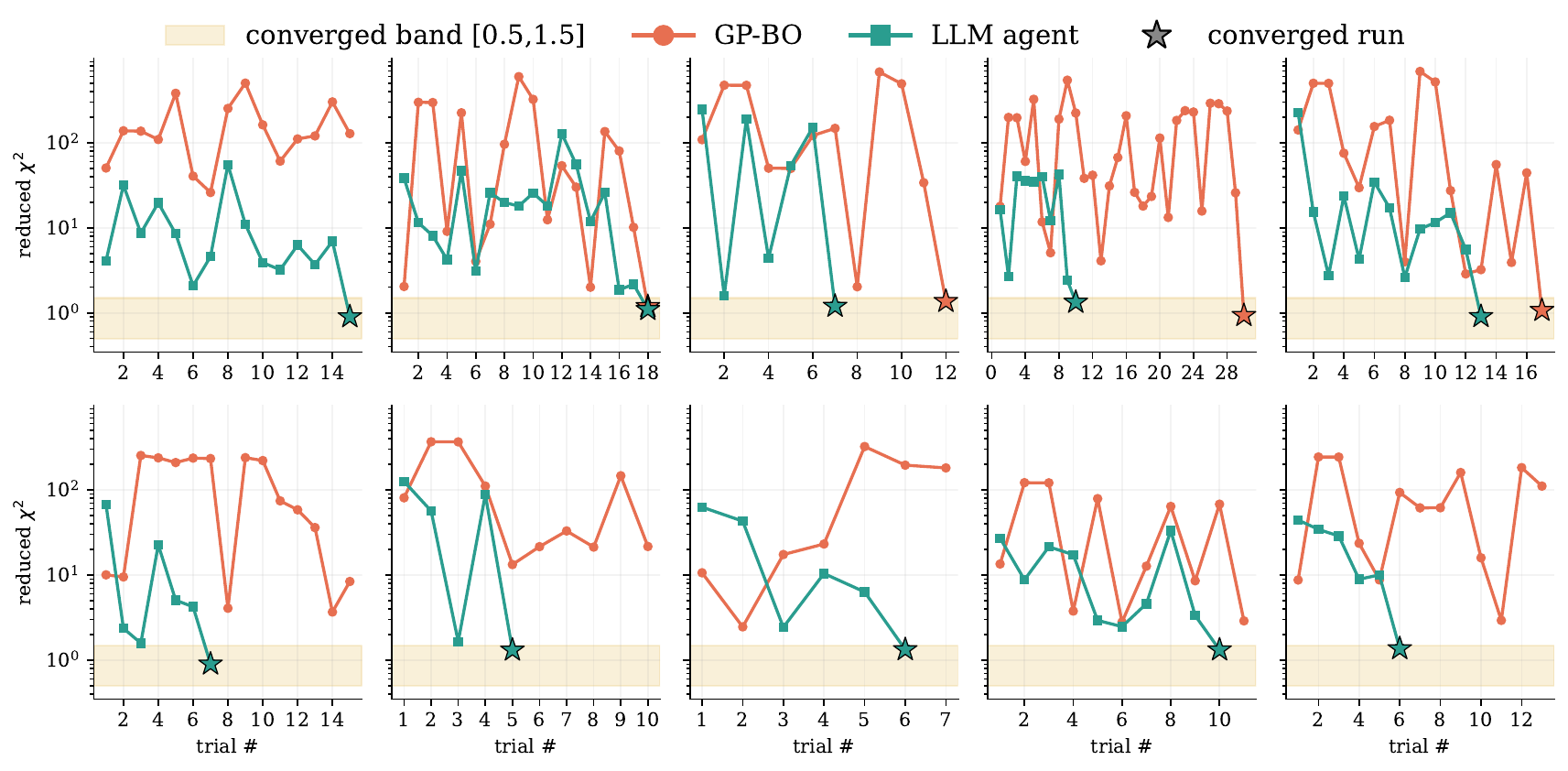}
    \caption{Reduced-$\tilde\chi^2$ vs. trial number for \AgenticTuner
  (teal) and BO baseline (orange). We present ten representative runs from our 20 experiments for legibility. To stress test our algorithm, we reduced the allowed CPU budget for both methods in the bottom row, which explains the total lack of convergence in BO runs.}
    \label{fig:convergence}
\end{figure}

\paragraph{Exploration.}
Let $t$ index expensive evaluations and $i$ index computationally free actions taken by the LLM agent in between evaluations.
Submitting $\theta_{t+1}$ to the simulator $\mathcal{M}$ is the only action that incurs cost $c(\theta_{t+1})$. Before doing so, the agent may choose to perform a number of cost-free observations $o_{t,i}$ and/or append to the workspace, $
    \left\{\Pfile,\ \Ccode, \,P^{\rm target}, \, \Guide,\, \Dhist,\ \Notebook\right\},
$
which functions as the working memory of the recently spawned agent.
The problem file $\mathcal{P}$ (see Appendix~\ref{app:prompt}) first orients the LLM by providing a natural-language description of the task, constraints, and a list of tunable parameters (and their allowed bounds). The LLM also has read-only access to the forward simulator's codebase $\mathcal{C}$ for further investigation.
Given the LLM's versatile ability to reason in natural language and adapt its strategies through in-context learning~\citep{brown2020languagemodelsfewshotlearners}, we expect it to match a human scientist's search performance when given the same domain knowledge about the problem and search space. This motivates the
guide $\mathcal{G}$ (see Appendix~\ref{app:doc}), which contains code documentation and distilled, `trial-and-error' intuitions human scientists rely on when fine-tuning simulations under a constrained budget~\citep{Yang_2020}. The agent may use the informal prior encoded in the guide to decide which regions of the parameter space to explore or exploit.
The trial history $\mathcal{D}_t$ and the lab notebook $\mathcal{N}_t^{(i)}$ are append-only. $\mathcal{D}_t$ is the accumulated log of past evaluations (parameter vector with simulated spectra), while $\mathcal{N}_t^{(i)}$ is the agent's commentary (e.g., reasoning, hypotheses, and lessons learned) written for future agent invocations. To avoid repeating previous failed attempts, the current agent may consult these records before making a decision.

\paragraph{Low-resolution probe.} Aside from inspecting the workspace, the agent can invoke a cheap, low-fidelity matter power spectrum simulator $\textsc{Probe}(\theta_{\rm cosmo})$, implemented using CAMB~\citep{Lewis_2000}, to obtain a dark-matter-only proxy for the power spectrum. The resulting power spectrum is an approximation in that $\textsc{Probe}$ only accepts cosmological parameters and thus neglects the effects of the remaining baryonic or subgrid parameters. We expect the agent to use the low-resolution proxy to narrow down the relevant cosmology before proceeding with the expensive high-resolution simulation.

\begin{figure}[b]
    \centering
    \includegraphics[width=1\linewidth]{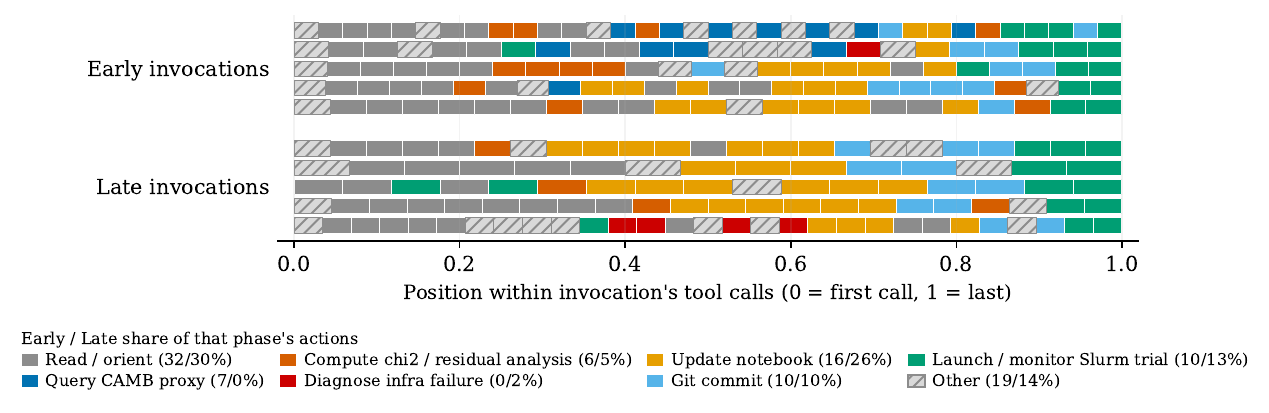}
\caption{Per-invocation timeline of \AgenticTuner's reasoning and tool calls, early vs. late in the run (pooled over independent runs; five shown per panel to save space). Actions are fetched and classified from the saved traces; \emph{Other} covers miscellaneous sanity-check code. Legend percentages (early/late) give each action's total share across \textit{all} runs.}
    \label{fig:reasoning}
\end{figure}

\section{Experiments and Results}

To compare \AgenticTuner and BO baseline in reaching convergence within a fixed budget, we run the experiment 20 times with different random seeds, true parameters, and simulator initial conditions.

\paragraph{Calibration performance.} Our results show that by incorporating the informal prior into the selection policy, \AgenticTuner achieves convergence using a median of 8.5 simulation calls (68\% CI: 6.0--14.1). In comparison, the simulation cost is nearly twice as low as BO, which requires a median of 17.5 calls (14.4--24.2).
Figure~\ref{fig:convergence} depicts the reduced $\tilde\chi^2$ trajectory for some example runs. We find that the trajectory of \AgenticTuner (teal) shows a consistent downward trend with occasional bumps, while the BO baseline (orange) fluctuates for many trials before occasionally dropping. Because BO's oscillatory trajectory persists well beyond the early exploration phase, we hypothesize that the LLM agent's advantage is due in part to its access to informative physical priors.

\paragraph{Anatomy of reasoning.} To gain insight into the LLM's decision-making process while tackling a long-horizon task, we examine the lab notebook and reasoning trace during the first and last half of each run's invocations (see Figure~\ref{fig:reasoning}). The LLM begins each new invocation with no memory other than what is written to disk, necessitating a fixed reorientation cost (about 30\% of the budget) by scanning the workspace files.
During the early stages, the agent focuses on calling cheap $\textsc{Probe}$ proxies (7\%) and doing residual analysis (6\%), initially to localize the cosmological parameters and later to narrow the sweep of each. It also checks for degeneracies between the parameters before submitting the expensive trial. After pinning the cosmological parameters, the agent stops relying on $\textsc{Probe}$ in later reasoning, as expected, and switches its token use to curate findings in the notebook $\Notebook$ and analyze the accumulating trial history $\Dhist$ (26\%). This shift in reasoning demonstrates that the LLM uses a hierarchical strategy that evolves based on past history and remaining budget.

\begin{figure}[t]
    \centering
    \includegraphics[width=\linewidth]{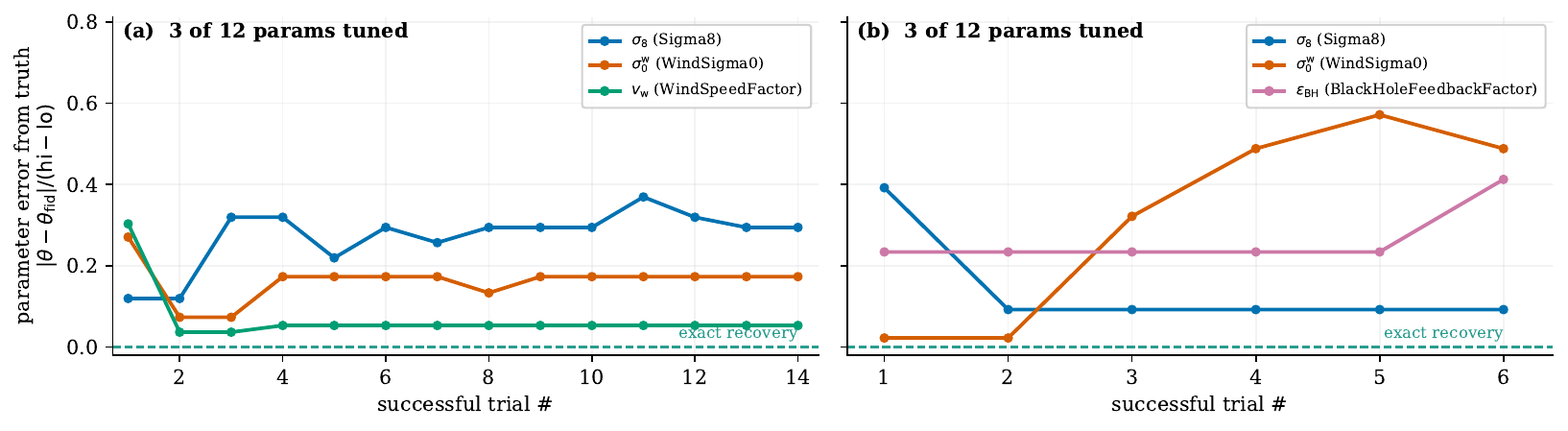}
    \caption{Per-trial parameter-recovery error $|\theta-\theta_{\rm fid}|/({\rm hi}-{\rm lo})$ across two representative converged \AgenticTuner runs; colored curves are the parameters the agent chose to tune.}
    \label{fig:recovery}
\end{figure}

\paragraph{Parameter recovery.} Across all runs, \AgenticTuner largely changes 2-5 parameters from their starting values, leaving the remaining parameters at the default values. This selective choice implies that the power spectra are constrained by a small set of parameters and that tuning all twelve parameters is not necessary to achieve convergence.
Figure~\ref{fig:recovery} shows the error in each parameter for two representative converged runs. Due to degeneracies in power spectra, different points in parameter space can map to the same observables~\citep{tirapongprasert2026degendetectorsymbolicrecoveryparameter}, which means that some parameters of the converged run may be further away from the ground truth than they started. In Panel A, the agent chooses to tune the wind feedback, whereas in Panel B, the agent achieves convergence by tuning $\sigma_8$ early.
We note that the observed divergences in parameter space due to degeneracies are not a limitation of \AgenticTuner, but rather a true physical degeneracy for this class of observable~\citep{vandaalen2011owls,amon2022nonlinear}.
Our guide (see Box~\ref{box:mp-gadget-diagnostics} in the Appendix) includes some of the well-known degeneracies and proposes corresponding recommendations, such as fixing $\Omega_m$ and using $\sigma_8$ as the main cosmology handle.

\section{Conclusion}
\label{sec:conclusion}

In this work, we present \AgenticTuner, an agentic calibration system that uses an LLM agent to propose the parameter point based on trial history and human-authored priors. We evaluate our system in calibrating a costly cosmological simulation and find that
\AgenticTuner outperforms the Bayesian optimization baseline in simulation cost across all runs.
Auditing the reasoning trace of the LLM agent demonstrates its ability to adapt the search strategy during a run from active tool use to synthesizing previous findings.
These results suggest that natural-language priors significantly improve the selection process.

Future work includes evaluating \AgenticTuner on a broader set of observables and ablating its context components (guide, codebase, low-resolution probe) to isolate what drives sample efficiency. It would also be useful to replicate our findings with open-weight LLMs, which we anticipate will achieve comparable calibration performance.
Finally, we plan to let the agent configure simulation volume and resolution for dynamic cost allocation and to benchmark its search strategy against a human scientist.

\section*{Acknowledgments}
C.~T. and M.~H. are supported in part by the Simons Collaboration on Learning the Universe. C.T. is supported by the Jocelyn Bell Burnell Research Fellowship. M.H. is supported by the Ashvin B. Chhabra and Daniela Bonafede-Chhabra Membership, the Ambrose Monell Foundation, and the IBM Einstein Fellowship. This project was developed in collaboration with the Keystone Astronomy and AI Workshop on Foundation Models \& Agentic Methods for Astrophysical Simulation at Carnegie Mellon University.

\newpage

\bibliographystyle{plainnat}
\bibliography{references}

@article{kugel2023flamingo,
  title   = {{FLAMINGO}: Calibrating large cosmological hydrodynamical simulations with machine learning},
  author  = {Kugel, Roi and Schaye, Joop and Schaller, Matthieu and Helly, John C. and Braspenning, Joey and Elbers, Willem and Frenk, Carlos S. and McCarthy, Ian G. and Kwan, Juliana and Salcido, Jaime and van Daalen, Marcel P. and Vandenbroucke, Bert and Bah{\'e}, Yannick M. and Borrow, Josh and Chaikin, Evgenii and Hu{\v{s}}ko, Filip and Jenkins, Adrian and Lacey, Cedric G. and Nobels, Folkert S. J. and Vernon, Ian},
  journal = {arXiv preprint arXiv:2306.05492},
  year    = {2023},
  note    = {Accepted for publication in MNRAS}
}

@article{hacc2026emulator,
  title   = {Emulator-Based Inference of Cosmological Subgrid Models},
  author  = {Ramachandra, Nesar and Frontiere, Nicholas and Buehlmann, Michael and Moran, Kelly R. and Emberson, J. D. and Heitmann, Katrin and Habib, Salman},
  journal = {arXiv preprint arXiv:2601.07306},
  year    = {2026}
}

@article{rogers2019bayesian,
  title   = {Bayesian emulator optimisation for cosmology: application to the Lyman-alpha forest},
  author  = {Rogers, Keir K. and Peiris, Hiranya V. and Pontzen, Andrew and Bird, Simeon and Verde, Licia and Font-Ribera, Andreu},
  journal = {Journal of Cosmology and Astroparticle Physics},
  volume  = {2019},
  number  = {02},
  pages   = {031},
  year    = {2019},
  month   = feb,
  publisher = {IOP Publishing},
  doi     = {10.1088/1475-7516/2019/02/031}
}

@article{zhang2023llmhpo,
  title   = {Using Large Language Models for Hyperparameter Optimization},
  author  = {Zhang, Michael R. and Desai, Nishkrit and Bae, Juhan and Lorraine, Jonathan and Ba, Jimmy},
  journal = {arXiv preprint arXiv:2312.04528},
  year    = {2023}
}

@article{madiraju2025optimindtune,
  title   = {OptiMindTune: A Multi-Agent Framework for Intelligent Hyperparameter Optimization},
  author  = {Madiraju, Meher Bhaskar and Madiraju, Meher Sai Preetam},
  journal = {arXiv preprint arXiv:2505.19205},
  year    = {2025}
}

@article{liu2024llambo,
  title   = {Large Language Models to Enhance Bayesian Optimization},
  author  = {Liu, Tennison and Astorga, Nicol{\'a}s and Seedat, Nabeel and van der Schaar, Mihaela},
  journal = {arXiv preprint arXiv:2402.03921},
  year    = {2024},
  note    = {LLAMBO; Poster, ICLR 2024}
}

@article{akiba2019optuna,
  title   = {Optuna: A Next-generation Hyperparameter Optimization Framework},
  author  = {Akiba, Takuya and Sano, Shotaro and Yanase, Toshihiko and Ohta, Takeru and Koyama, Masanori},
  journal = {arXiv preprint arXiv:1907.10902},
  year    = {2019},
  note    = {Applied Data Science track, KDD 2019}
}

@incollection{jette2023slurm,
  author    = {Jette, M. A. and Wickberg, T.},
  title     = {Architecture of the {Slurm} Workload Manager},
  booktitle = {Job Scheduling Strategies for Parallel Processing},
  editor    = {Klus{\'a}{\v{c}}ek, D. and Corbal{\'a}n, J. and Rodrigo, G. P.},
  series    = {Lecture Notes in Computer Science},
  volume    = {14283},
  publisher = {Springer},
  address   = {Cham},
  year      = {2023},
  doi       = {10.1007/978-3-031-43943-8_1}
}

@article{bird2022astridI,
  title   = {The {ASTRID} simulation: galaxy formation and reionization},
  author  = {Bird, Simeon and Ni, Yueying and Di Matteo, Tiziana and Croft, Rupert and Feng, Yu and Chen, Nianyi},
  journal = {Monthly Notices of the Royal Astronomical Society},
  volume  = {512},
  number  = {3},
  pages   = {3703--3716},
  year    = {2022},
  month   = mar,
  publisher = {Oxford University Press},
  doi     = {10.1093/mnras/stac648}
}

@article{ni2022astridII,
  title   = {The {ASTRID} simulation: the evolution of supermassive black holes},
  author  = {Ni, Yueying and Di Matteo, Tiziana and Bird, Simeon and Croft, Rupert and Feng, Yu and Chen, Nianyi and Tremmel, Michael and DeGraf, Colin and Li, Yin},
  journal = {Monthly Notices of the Royal Astronomical Society},
  volume  = {513},
  number  = {1},
  pages   = {670--692},
  year    = {2022},
  month   = feb,
  publisher = {Oxford University Press},
  doi     = {10.1093/mnras/stac351}
}

@book{rasmussen2006gaussian,
  title     = {Gaussian Processes for Machine Learning},
  author    = {Rasmussen, Carl Edward and Williams, Christopher K. I.},
  publisher = {MIT Press},
  year      = {2006},
  address   = {Cambridge, MA},
  isbn      = {0-262-18253-X}
}

@misc{dongre2025drift,
  title         = {Drift No More? Context Equilibria in Multi-Turn LLM Interactions},
  author        = {Dongre, Vardhan and Rossi, Ryan A. and Lai, Viet Dac and Yoon, David Seunghyun and Hakkani-T{\"u}r, Dilek and Bui, Trung},
  year          = {2025},
  eprint        = {2510.07777},
  archivePrefix = {arXiv}
}

@misc{xia2026diagnosing,
  title         = {Diagnosing and Mitigating Context Rot in Long-horizon Search},
  author        = {Xia, Shijie and Wang, Yikun and Huang, Zhen and Liu, Pengfei},
  year          = {2026},
  eprint        = {2606.29718},
  archivePrefix = {arXiv}
}

@misc{wang2026horizon,
  title         = {The Long-Horizon Task Mirage? Diagnosing Where and Why Agentic Systems Break},
  author        = {Wang, Xinyu Jessica and Bai, Haoyue and Sun, Yiyou and Wang, Haorui and Zhang, Shuibai and Hu, Wenjie and Schroder, Mya and Mutlu, Bilge and Song, Dawn and Nowak, Robert D.},
  year          = {2026},
  eprint        = {2604.11978},
  archivePrefix = {arXiv}
}

@article{Lewis_2000,
   title={Efficient Computation of Cosmic Microwave Background Anisotropies in Closed Friedmann‐Robertson‐Walker Models},
   volume={538},
   ISSN={1538-4357},
   url={http://dx.doi.org/10.1086/309179},
   DOI={10.1086/309179},
   number={2},
   journal={The Astrophysical Journal},
   publisher={American Astronomical Society},
   author={Lewis, Antony and Challinor, Anthony and Lasenby, Anthony},
   year={2000},
   month=Aug, pages={473–476} }

@misc{brown2020languagemodelsfewshotlearners,
      title={Language Models are Few-Shot Learners}, 
      author={Tom B. Brown and Benjamin Mann and Nick Ryder and Melanie Subbiah and Jared Kaplan and Prafulla Dhariwal and Arvind Neelakantan and Pranav Shyam and Girish Sastry and Amanda Askell and Sandhini Agarwal and Ariel Herbert-Voss and Gretchen Krueger and Tom Henighan and Rewon Child and Aditya Ramesh and Daniel M. Ziegler and Jeffrey Wu and Clemens Winter and Christopher Hesse and Mark Chen and Eric Sigler and Mateusz Litwin and Scott Gray and Benjamin Chess and Jack Clark and Christopher Berner and Sam McCandlish and Alec Radford and Ilya Sutskever and Dario Amodei},
      year={2020},
      eprint={2005.14165},
      archivePrefix={arXiv},
      primaryClass={cs.CL},
      url={https://arxiv.org/abs/2005.14165}, 
}

@misc{anthropic2026sonnet5,
  author       = {{Anthropic}},
  title        = {Claude Sonnet 5},
  howpublished = {\url{https://platform.claude.com/docs/en/models/sonnet-5/overview}},
  year         = {2026},
}

@article{Yang_2020,
   title={On hyperparameter optimization of machine learning algorithms: Theory and practice},
   volume={415},
   ISSN={0925-2312},
   url={http://dx.doi.org/10.1016/j.neucom.2020.07.061},
   DOI={10.1016/j.neucom.2020.07.061},
   journal={Neurocomputing},
   publisher={Elsevier BV},
   author={Yang, Li and Shami, Abdallah},
   year={2020},
   month=Nov, pages={295–316} }

@misc{anthropic2025context,
  author       = {{Anthropic}},
  title        = {Effective Context Engineering for {AI} Agents},
  year         = {2025},
  month        = sep,
  howpublished = {Anthropic Engineering Blog},
  url          = {https://www.anthropic.com/engineering/effective-context-engineering-for-ai-agents}
}

@inproceedings{hvarfner2022pibo,
  title     = {{$\pi$BO}: Augmenting Acquisition Functions with User Beliefs for {B}ayesian Optimization},
  author    = {Hvarfner, Carl and Stoll, Danny and Souza, Artur and Lindauer, Marius and Hutter, Frank and Nardi, Luigi},
  booktitle = {International Conference on Learning Representations (ICLR)},
  year      = {2022},
  eprint        = {2204.11051},
  archivePrefix = {arXiv}
}

@inproceedings{feurer2015initializing,
  title     = {Initializing {B}ayesian Hyperparameter Optimization via Meta-Learning},
  author    = {Feurer, Matthias and Springenberg, Jost Tobias and Hutter, Frank},
  booktitle = {Proceedings of the AAAI Conference on Artificial Intelligence},
  volume    = {29},
  pages     = {1128--1135},
  year      = {2015},
  doi       = {10.1609/aaai.v29i1.9354}
}

@article{bergstra2011algorithms,
  title={Algorithms for hyper-parameter optimization},
  author={Bergstra, James and Bardenet, R{\'e}mi and Bengio, Yoshua and K{\'e}gl, Bal{\'a}zs},
  journal={Advances in neural information processing systems},
  volume={24},
  year={2011}
}

@article{lindauer2022smac3,
  title={SMAC3: A versatile Bayesian optimization package for hyperparameter optimization},
  author={Lindauer, Marius and Eggensperger, Katharina and Feurer, Matthias and Biedenkapp, Andr{\'e} and Deng, Difan and Benjamins, Carolin and Ruhkopf, Tim and Sass, Ren{\'e} and Hutter, Frank},
  journal={Journal of Machine Learning Research},
  volume={23},
  number={54},
  pages={1--9},
  year={2022}
}

@article{villaescusa2021camels,
  title={The CAMELS project: Cosmology and astrophysics with machine-learning simulations},
  author={Villaescusa-Navarro, Francisco and Angl{\'e}s-Alc{\'a}zar, Daniel and Genel, Shy and Spergel, David N and S. Somerville, Rachel and Dave, Romeel and Pillepich, Annalisa and Hernquist, Lars and Nelson, Dylan and Torrey, Paul and others},
  journal={The Astrophysical Journal},
  volume={915},
  number={1},
  pages={71},
  year={2021},
  publisher={The American Astronomical Society}
}

@article{rose2025introducing,
  title={Introducing the DREAMS project: DaRk mattEr and astrophysics with machine learning and simulations},
  author={Rose, Jonah C and Torrey, Paul and Villaescusa-Navarro, Francisco and Lisanti, Mariangela and Nguyen, Tri and Roy, Sandip and Kollmann, Kassidy E and Vogelsberger, Mark and Cyr-Racine, Francis-Yan and Medvedev, Mikhail V and others},
  journal={The Astrophysical Journal},
  volume={982},
  number={2},
  pages={68},
  year={2025},
  publisher={The American Astronomical Society}
}

@article{vandaalen2011owls,
  title={The effects of galaxy formation on the matter power spectrum: a challenge for precision cosmology},
  author={van Daalen, Marcel P. and Schaye, Joop and Booth, C. M. and Dalla Vecchia, Claudio},
  journal={Monthly Notices of the Royal Astronomical Society},
  volume={415},
  number={4},
  pages={3649--3665},
  year={2011},
  eprint={1104.1174},
  archivePrefix={arXiv},
  publisher={Oxford University Press}
}

@misc{tirapongprasert2026degendetectorsymbolicrecoveryparameter,
      title={DegenDetector: Symbolic Recovery of Parameter Degeneracies in Bayesian Posteriors}, 
      author={Chaipat Tirapongprasert and Matthew Ho},
      year={2026},
      eprint={2607.08755},
      archivePrefix={arXiv},
      primaryClass={astro-ph.IM},
      url={https://arxiv.org/abs/2607.08755}, 
}

@article{amon2022nonlinear,
  title={A non-linear solution to the $S_8$ tension?},
  author={Amon, Alexandra and Efstathiou, George},
  journal={Monthly Notices of the Royal Astronomical Society},
  volume={516},
  number={4},
  pages={5355--5366},
  year={2022},
  eprint={2206.11794},
  archivePrefix={arXiv},
  publisher={Oxford University Press}
}

@article{leclercq2018bayesian,
  title={Bayesian optimization for likelihood-free cosmological inference},
  author={Leclercq, Florent},
  journal={Physical Review D},
  volume={98},
  number={6},
  pages={063511},
  year={2018},
  publisher={APS}
}
\newpage
\appendix
\section{Simulation and Discrepancy Metric Details}
\label{app:sim-details}

\paragraph{Simulation settings.} We use $N_{\rm grid}=128$ (i.e., $128^3$ particles) and $L_{\rm box}=100$ Mpc/h, matching the target box specified in the task prompt (Appendix~\ref{app:prompt}); box size sets the simulated volume and largest resolvable scale, grid resolution sets the smallest resolvable scale. All runs were submitted as Slurm jobs~\citep{jette2023slurm}, each with 32 CPU cores.

\paragraph{Discrepancy metric.} $\sigma_i^2 = 2 (P_i^{\rm target})^2/N_i$ is the cosmic variance of a power spectrum estimator averaged over $N_i$ independent Fourier modes. We drop bins below the fundamental mode $k_{\min} = 2\pi/L_{\rm box}$ and above the grid's Nyquist frequency $k_{\max} = \pi\,N_{\rm grid}/L_{\rm box}$.

\paragraph{BO baseline configuration.} Our BO baseline employs Optuna's \texttt{GPSampler}~\citep{akiba2019optuna} to minimize $\chi^2$ across the 12-dimensional box $\Theta$. The optimization process begins with five rounds of random exploration, after which the GP surrogate handles the selection. Failed trials receive a large finite penalty ($\chi^2 = 10^7$) instead of being pruned, ensuring the surrogate remains well-defined when trials are scarce. On each restart, the full trial history $\mathcal{D}_t$ is replayed into a new study rather than resumed incrementally. The search ends when $\tilde\chi^2 < 1$ or the CPU-hour budget is exhausted.

\section{Task Description Prompts}\label{app:prompt}

We have included the full user prompt below for reproducibility purposes. All LLM calls use Claude Sonnet 5~\citep{anthropic2026sonnet5}. A single calibration run fits within the monthly usage allowance of a \$$20$ Claude Pro plan, making the approach accessible without metered API billing.

\subsection{Problem Statement}

Each trial is executed by a new agent instance with no conversational memory. It maintains continuity by reading and writing files (\texttt{lab\_notebook.md}, \texttt{trial\_state.yaml}). The first section of the prompt introduces the agent to the task and the workspace.

\refstepcounter{promptboxctr}
\begin{promptbox}{Task framing}
\label{box:game-prompt-framing}
\par\medskip\noindent\pbbold{Your task: calibrate MP-Gadget parameters against a target power spectrum}\par\smallskip
You are one instance in a series -- fresh agent invoked each trial, no memory except what's written to disk. Read, in order, each session:

\begin{enumerate}[leftmargin=*,itemsep=1pt,topsep=2pt,parsep=0pt]
\item \texttt{lab\_\allowbreak{}notebook.\allowbreak{}md} -- past reasoning and what you've learned so far.
\item \texttt{trial\_\allowbreak{}state.\allowbreak{}yaml} -- structured record of every trial run so far (params, chi2, cpu\_hours), starting guess, CPU-hour budget.
\item \texttt{.\allowbreak{}.\allowbreak{}/\allowbreak{}.\allowbreak{}.\allowbreak{}/\allowbreak{}target/\allowbreak{}target\_\allowbreak{}powerspectrum.\allowbreak{}csv} -- power spectrum you're trying to match, and \texttt{.\allowbreak{}.\allowbreak{}/\allowbreak{}.\allowbreak{}.\allowbreak{}/\allowbreak{}target/\allowbreak{}README.\allowbreak{}md} for column definitions. (Two levels up from this workspace, alongside a \texttt{judge/\allowbreak{}} directory you should never read -- see Boundaries below. In practice \texttt{judge/\allowbreak{}} won't exist from where you're running: hidden from this session's filesystem view.)
\end{enumerate}

If \texttt{trial\_\allowbreak{}state.\allowbreak{}yaml} has zero trials, starting guess is the \texttt{starting\_\allowbreak{}guess} block already in that file -- run it first as trial 1 to establish baseline before changing anything.

\end{promptbox}

\refstepcounter{promptboxctr}
\begin{promptbox}{Simulation box specification}
\par\medskip\noindent\pbbold{The simulation box}\par\smallskip
Target produced by a box with L=100 Mpc/h, N=128, final snapshot z=0.

Every real MP-Gadget trial uses its own independent random seed (not shared with target's), so chi2 will be noisier trial to trial even at fixed parameters -- expected, not a bug. See \texttt{docs/\allowbreak{}sim-\allowbreak{}guide/\allowbreak{}mp-\allowbreak{}gadget.\allowbreak{}md}'s \texttt{Seed} entry for what this noise should look like.

\end{promptbox}

\subsection{Low-fidelity CAMB pre-screening}

In our setup, full MP-Gadget trials cost CPU hours, while a halofit-based nonlinear power spectrum is considered free.
Before allocating funds for the actual run, the agent is encouraged to narrow the five cosmological parameters at a low cost by exposing this as an explicit, unmetered tool.

\refstepcounter{promptboxctr}
\begin{promptbox}{Free CAMB proxy stage}
\par\medskip\noindent\pbbold{Low-fidelity CAMB stage (free -- optional, use whenever useful)}\par\smallskip
\texttt{src/\allowbreak{}camb\_\allowbreak{}proxy.\allowbreak{}py}'s \texttt{nonlinear\_\allowbreak{}matter\_\allowbreak{}power()} computes CAMB's halofit nonlinear matter power spectrum for a given cosmology, in seconds, no Slurm job, no CPU-hour cost:

\begin{Verbatim}[breaklines=true,breakanywhere=true,fontsize=\footnotesize]
import sys
sys.path.insert(0, "{{PROJECT_ROOT}}/src")
import camb_proxy

pk = camb_proxy.nonlinear_matter_power(
    k_h_per_mpc=target_k,
    omega0=..., omega_baryon=..., sigma8=..., hubble_param=..., primordial_index=...,
    z=0.0,
)
\end{Verbatim}
\pbbold{Only helps with the 5 cosmological parameters} (\texttt{Omega0}, \texttt{OmegaBaryon}, \texttt{Sigma8}, \texttt{PrimordialIndex}, \texttt{HubbleParam}). CAMB/halofit has no representation of AGN feedback or galactic winds -- dark-matter-only calculation. Won't respond to \texttt{WindSigma0}, \texttt{WindSpeedFactor}, \texttt{WindFreeTravelLength}, \texttt{WindFreeTravelDensFac}, \texttt{BlackHoleFeedbackFactor}, \texttt{BlackHoleAccretionFactor}, or \texttt{SeedBlackHoleMass} -- don't waste queries tuning those. Expect CAMB to track target at low-k end of this experiment's range (k close to box's fundamental mode, \textasciitilde{}0.0628 h/Mpc), increasingly diverging toward k=4.02 h/Mpc (box's particle Nyquist, top of scored range), since true target includes baryonic suppression/feedback signatures CAMB can't capture -- divergence itself is informative (tells you something about the baryonic physics real trials must pin down), not a sign CAMB is broken. \pbbold{Note}: a free CAMB-only sweep on this experiment family's earlier (smaller) box found \texttt{Sigma8} alone can fake much of that suppression by sitting below true value across this k-range -- don't treat a CAMB-only match as confirmation cosmology is exactly right; real HR trials (which include baryons) are the only real check.

Sensible strategy: use CAMB queries (as many as wanted, free) to find a cosmology matching target well at low-to-mid k, before spending real HR trials. Then use real trials to refine that cosmology and fit the 7 baryonic parameters, which only a real simulation can constrain.

\end{promptbox}

\subsection{Parameter space}

\begin{table}[h]
\centering
\small
\begin{tabular}{llc}
\hline
\textbf{Parameter} & \textbf{Role} & \textbf{Prior box} \\
\hline
$\Omega_m$ (\texttt{Omega0}) & cosmological & $[0.20, 0.45]$ \\
$\Omega_b$ (\texttt{OmegaBaryon}) & cosmological & $[0.03, 0.06]$ \\
$\sigma_8$ (\texttt{Sigma8}) & cosmological & $[0.60, 1.00]$ \\
$n_s$ (\texttt{PrimordialIndex}) & cosmological & $[0.90, 1.05]$ \\
$h$ (\texttt{HubbleParam}) & cosmological & $[0.60, 0.80]$ \\
\texttt{WindSigma0} & wind mass-loading & $[200, 500]$ km/s \\
\texttt{WindSpeedFactor} & wind kick speed & $[1.5, 6.0]$ \\
\texttt{WindFreeTravelLength} & wind decoupling & $[5, 60]$ \\
\texttt{WindFreeTravelDensFac} & wind decoupling & $[0.02, 0.5]$ \\
\texttt{BlackHoleFeedbackFactor} & AGN feedback & $[0.01, 0.15]$ \\
\texttt{BlackHoleAccretionFactor} & AGN accretion & $[20, 300]$ \\
\texttt{SeedBlackHoleMass} & AGN seeding & $[5\times10^{-6}, 10^{-4}]$ \\
\hline
\end{tabular}
\caption{The twelve-parameter search box $\Theta$.}
\label{tab:params}
\end{table}

This section provides a list of tunable parameters, including five cosmological parameters that a linear/halofit calculation can reasonably constrain and the remainder seven subgrid feedback parameters.

\refstepcounter{promptboxctr}
\begin{promptbox}{Tunable parameters}
\par\medskip\noindent\pbbold{What you're tuning (in the real HR simulation)}\par\smallskip
Twelve MP-Gadget parameters:

\begin{itemize}[leftmargin=*,itemsep=1pt,topsep=2pt,parsep=0pt]
\item Cosmological: \texttt{Omega0}, \texttt{OmegaBaryon}, \texttt{Sigma8}, \texttt{PrimordialIndex}, \texttt{HubbleParam}.
\item Subgrid stellar/AGN feedback: \texttt{WindSigma0}, \texttt{WindSpeedFactor}, \texttt{WindFreeTravelLength}, \texttt{WindFreeTravelDensFac}, \texttt{BlackHoleFeedbackFactor}, \texttt{BlackHoleAccretionFactor}, \texttt{SeedBlackHoleMass}.
\end{itemize}

See "Available documentation" below for \texttt{docs/\allowbreak{}sim-\allowbreak{}guide/\allowbreak{}mp-\allowbreak{}gadget.\allowbreak{}md} (read first for these mechanisms) and the MP-Gadget source fallback.

\pbbold{Important}: \texttt{run\_\allowbreak{}trial()}'s \texttt{params} dict only accepts the twelve names above; anything else passed is silently ignored (never written to paramfiles).

\end{promptbox}

\subsection{Trial execution protocol}

Because the agent is to submit a multi-hour Slurm job on a shared node, we provide boilerplate code that handles the minutia of such orchestration.

\refstepcounter{promptboxctr}
\begin{promptbox}{Launching and monitoring a real HR trial}
\par\medskip\noindent\pbbold{How to run a real HR trial}\par\smallskip
Trial submission-and-wait handled by shared script \texttt{src/\allowbreak{}run\_\allowbreak{}trial\_\allowbreak{}daemon.\allowbreak{}py} -- don't write your own \texttt{subprocess}/\texttt{nohup} submission code.

\pbbold{Step 1: check whether a trial is already in flight.} Before anything else, read \texttt{trial.\allowbreak{}status} in this workspace if it exists:

\begin{Verbatim}[breaklines=true,breakanywhere=true,fontsize=\footnotesize]
import json
from pathlib import Path
status = json.loads(Path("trial.status").read_text()) if Path("trial.status").exists() else None
\end{Verbatim}
\begin{itemize}[leftmargin=*,itemsep=1pt,topsep=2pt,parsep=0pt]
\item Missing, or \texttt{state} is \texttt{"done"}/\texttt{"failed"} \pbbold{and} \texttt{trial\_\allowbreak{}id} already recorded in \texttt{trial\_\allowbreak{}state.\allowbreak{}yaml}'s \texttt{trials} list -- nothing in flight, nothing pending. Go to Step 2 to plan and launch new trial (after any free tool queries wanted first).
\item \texttt{state} is \texttt{"done"} or \texttt{"failed"} but \texttt{trial\_\allowbreak{}id} NOT yet in \texttt{trial\_\allowbreak{}state.\allowbreak{}yaml} -- trial just finished, needs recording. Read \texttt{trial.\allowbreak{}result.\allowbreak{}json} (chi2, reduced\_chi2, per-bin \texttt{table}, cpu\_hours, wall\_time\_min) and follow "What to do with the result" below before considering new trial.
\item \texttt{state} is \texttt{"writing\_\allowbreak{}paramfiles"}, \texttt{"submitting"}, or \texttt{"running"} -- submission genuinely in flight (may still be retrying \texttt{sbatch} under Slurm QOS contention with a sibling condition -- expected, not a bug). \pbbold{Don't launch another.} Note this briefly in \texttt{lab\_\allowbreak{}notebook.\allowbreak{}md} (optionally use any free tools available while you wait), then exit -- driver won't spawn next instance until this one's lock clears.
\end{itemize}

\pbbold{Step 2: launch a new trial.} Decide twelve parameters as before, write reasoning to \texttt{lab\_\allowbreak{}notebook.\allowbreak{}md}'s "Next Planned Trial" section (see step 1 under "What to do with the result" below) BEFORE launching, then:

\begin{Verbatim}[breaklines=true,breakanywhere=true,fontsize=\footnotesize]
import json
from pathlib import Path
next_trial = {
    "trial_id": next_id,  # len(trial_state.yaml's trials) + 1
    "params": {"Omega0": ..., "OmegaBaryon": ..., "Sigma8": ...,
               "PrimordialIndex": ..., "HubbleParam": ...,
               "WindSigma0": ..., "WindSpeedFactor": ...,
               "WindFreeTravelLength": ..., "WindFreeTravelDensFac": ...,
               "BlackHoleFeedbackFactor": ..., "BlackHoleAccretionFactor": ...,
               "SeedBlackHoleMass": ...},
    "seed": trial_seed,  # random.randint(1, 2**31 - 1) -- fresh every trial, never reused
    "job_name": f"calib-h6t1-llmcamb-{next_id}",
    "sigma_mode": "unpaired",
}
Path("next_trial.json").write_text(json.dumps(next_trial))
\end{Verbatim}
Then, from a Bash tool call:

\begin{Verbatim}[breaklines=true,breakanywhere=true,fontsize=\footnotesize]
nohup python3 {{PROJECT_ROOT}}/src/run_trial_daemon.py \
    --request next_trial.json \
    --target ../../target/target_powerspectrum.csv \
    --box-config ../../box_config/hr_box_config.py:HR_BOX \
    > trial_daemon.log 2>&1 &
disown
\end{Verbatim}
Returns almost instantly -- daemon does the waiting, detached from your session. Don't bump your own tool timeout to cover full trial duration and wait on it inline; just confirm it launched (e.g. \texttt{trial.\allowbreak{}status} shows \texttt{"writing\_\allowbreak{}paramfiles"} or \texttt{"submitting"} a moment later) and exit. It handles submission retries itself, writes \texttt{trial.\allowbreak{}lock} / \texttt{trial.\allowbreak{}status} / \texttt{trial.\allowbreak{}result.\allowbreak{}json} for the next instance to read.

\end{promptbox}

\subsection{Recording results and convergence criteria}

To avoid post-hoc rationalization, the agent must write their predicted reasoning before seeing the trial outcome. The two hard stopping conditions are a reduced-$\chi^2 < 1$ convergence criterion and the CPU-hour budget cap.

\refstepcounter{promptboxctr}
\begin{promptbox}{Recording, updating, and convergence}
\par\medskip\noindent\pbbold{What to do with the result}\par\smallskip
\begin{enumerate}[leftmargin=*,itemsep=1pt,topsep=2pt,parsep=0pt]
\item \pbbold{Before} writing \texttt{next\_\allowbreak{}trial.\allowbreak{}json} and launching the daemon, write reasoning for the parameters about to try into \texttt{lab\_\allowbreak{}notebook.\allowbreak{}md} under "Next Planned Trial" before seeing the outcome. Hard project convention: never write reasoning after seeing the result and call it a prediction. Free tool queries (any available above) don't need this pre-registration (free/reversible/no experimental integrity concern), but log what you learned from them in \texttt{lab\_\allowbreak{}notebook.\allowbreak{}md} too, so the next fresh instance can pick up where you left off.
\item \pbbold{When you find} \texttt{trial.\allowbreak{}status} shows \texttt{"done"} or \texttt{"failed"} for a \texttt{trial\_\allowbreak{}id} not yet in \texttt{trial\_\allowbreak{}state.\allowbreak{}yaml} (see Step 1 of "How to run a real HR trial" above), append a dated entry to the "Trial-by-trial log" section of \texttt{lab\_\allowbreak{}notebook.\allowbreak{}md}: what you tried, why, what happened, what it tells you (failed/crashed trial informative too -- note what parameter region to avoid and why, don't just discard it). If a result looks like an outlier relative to recent trend, consider whether trial-to-trial seed noise explains it before concluding your parameters moved things the wrong direction.
\item Append a structured record to \texttt{trial\_\allowbreak{}state.\allowbreak{}yaml}'s \texttt{trials} list with the same fields shown in its commented example -- including \texttt{seed} (read \texttt{params}/\texttt{seed} back from \texttt{next\_\allowbreak{}trial.\allowbreak{}json} or the record already written to \texttt{lab\_\allowbreak{}notebook.\allowbreak{}md}, and \texttt{chi2}/\texttt{reduced\_\allowbreak{}chi2}/\texttt{cpu\_\allowbreak{}hours}/\texttt{wall\_\allowbreak{}time\_\allowbreak{}min} from \texttt{trial.\allowbreak{}result.\allowbreak{}json}). Free tool queries aren't real trials and don't get a \texttt{trial\_\allowbreak{}state.\allowbreak{}yaml} entry or budget charge -- just log them in the notebook.
\item Update \texttt{lab\_\allowbreak{}notebook.\allowbreak{}md}'s "Current Understanding" and "Lessons and Constraints" sections if this trial (or a free tool query) changed your model of the parameter -> P(k) mapping. Prune stale reasoning rather than letting this section grow forever.
\item Decide: if reduced\_chi2 (from a real HR trial) is below 1 -- i.e. the simulated spectrum sits within the cosmic-variance noise floor of the target -- across the k-range [k\_f\textasciitilde{}=0.0628, 4.02] h/Mpc, you've converged -- set \texttt{converged: true} in \texttt{trial\_\allowbreak{}state.\allowbreak{}yaml} and stop. A reduced\_chi2 at or above 1 means the spectrum still departs from target by more than the expected noise -- treat it as a real mismatch to keep reducing. If \texttt{budget\_\allowbreak{}cpu\_\allowbreak{}hours\_\allowbreak{}used} is at or past \texttt{budget\_\allowbreak{}cpu\_\allowbreak{}hours\_\allowbreak{}cap}, also stop (note in the notebook that you ran out of budget without converging -- a valid, informative outcome, not a failure to hide).
\item Otherwise, launch exactly one new real HR trial (free tool queries don't count toward this limit -- use them as many times as useful), then exit. At most one trial may ever be in flight at a time -- \texttt{run\_\allowbreak{}trial\_\allowbreak{}daemon.\allowbreak{}py} enforces this itself (it refuses to start a second submission while \texttt{trial.\allowbreak{}lock} names a live process), so don't try to work around it. This keeps each invocation short-lived and robust to crashes/disconnects -- the point is never more than one trial in flight, not more than one invocation per trial (a single trial's submission-and-wait may span several fresh instances if retrying under QOS contention, which is fine).
\end{enumerate}

\end{promptbox}

\subsection{Boundaries against finding the answer key}

An unconstrained agent could bypass the need to infer the ground-truth parameters from simulation output and instead search for them directly on the filesystem. This section closes the loophole procedurally by instructing the agent not to waste extra tokens trying to look for the answer. Our harness also enforces the constraint structurally by excluding the \texttt{judge/} directory from the session's sandbox view.

\refstepcounter{promptboxctr}
\begin{promptbox}{Boundaries}
\par\medskip\noindent\pbbold{Boundaries}\par\smallskip
Stay within this workspace, the target/ directory two levels up, and the MP-Gadget source/docs. Don't look for the true parameter values elsewhere on the filesystem -- point of this experiment is whether you can find them through physical reasoning and iterative simulation, not by searching for the answer. In practice a \texttt{judge/\allowbreak{}} directory that would contain them is hidden from this session entirely, so searching for it is also wasted effort.

\end{promptbox}

\subsection{Available documentation}

\refstepcounter{promptboxctr}
\begin{promptbox}{Documentation pointers}
\par\medskip\noindent\pbbold{Available documentation}\par\smallskip
Read \texttt{docs/\allowbreak{}sim-\allowbreak{}guide/\allowbreak{}mp-\allowbreak{}gadget.\allowbreak{}md} first -- single dense reference covering how each MP-Gadget parameter affects the matter power spectrum, plus a diagnostics guide mapping observed P(k) discrepancies back to which parameter(s) to try and in which direction. Fall back to the actual MP-Gadget source (path given by \texttt{MPGADGET\_\allowbreak{}ROOT}, see \texttt{src/\allowbreak{}calibration\_\allowbreak{}env.\allowbreak{}py}; inline docs in \texttt{gadget/\allowbreak{}params.\allowbreak{}c}/\texttt{genic/\allowbreak{}params.\allowbreak{}c}, \texttt{README.\allowbreak{}rst}) for anything not covered in enough depth, plus whatever you already know about cosmological simulations from training.

\end{promptbox}

\section{Guide and documentation}\label{app:doc}

\refstepcounter{promptboxctr}
\begin{promptbox}{Heuristic guide}
\label{box:mp-gadget-diagnostics}
\par\medskip\noindent\pbbold{Diagnostics Guide: Parameter -> Power Spectrum Symptom Mapping}\par\smallskip
Use this to select which parameter(s) to perturb based on observed discrepancy between simulated and target P(k). Every parameter named here is one of the 12 you can tune (see TASK.md).

\par\medskip\noindent\pbbold{If overall amplitude is too low/high (roughly scale-independent across all k)}\par\smallskip
\begin{itemize}[leftmargin=*,itemsep=1pt,topsep=2pt,parsep=0pt]
\item \pbbold{Increase/decrease \texttt{Sigma8}} -- direct, clean, monotonic amplitude lever in this setup (see \texttt{PrimordialAmp} note above for why \texttt{PrimordialAmp} itself isn't useful here). Preferred first choice.
\item \pbbold{Increase/decrease \texttt{Omega0}} -- also boosts amplitude but couples to growth rate and shifts turnover scale; use only if \texttt{Sigma8} alone can't explain the discrepancy (e.g., background expansion history also off).
\item These two are partially degenerate at fixed z; prefer \texttt{Sigma8} for a pure amplitude fix, reserve \texttt{Omega0} for cases where shape near turnover is also off.
\end{itemize}

\par\medskip\noindent\pbbold{If large-scale slope/tilt is wrong (power ratio between low-k and high-k off, but amplitude near pivot k\textasciitilde{}0.05 Mpc\textasciicircum{}-1 is fine)}\par\smallskip
\begin{itemize}[leftmargin=*,itemsep=1pt,topsep=2pt,parsep=0pt]
\item \pbbold{Adjust \texttt{PrimordialIndex}} -- increasing raises relative power at high k vs low k (blue tilt), decreasing does the opposite (red tilt). Nearly scale-independent within the linear regime, degenerate with \texttt{Sigma8} only in overall normalization, not slope. Monotonic.
\end{itemize}

\par\medskip\noindent\pbbold{If small-scale power (k >\textasciitilde{} 1 h/Mpc, but still below this setup's k\_Nyquist \textasciitilde{} 4 h/Mpc -- see Ngrid above) is suppressed relative to target, broad trough shape}\par\smallskip
\begin{itemize}[leftmargin=*,itemsep=1pt,topsep=2pt,parsep=0pt]
\item \pbbold{Rule out resolution first}: \texttt{Ngrid}/\texttt{BoxSize} are fixed here, so a suppression that looks purely numerical (rather than shaped like feedback) is worth flagging in your notebook, but isn't something you can resolve by changing resolution.
\item \pbbold{Try wind feedback first}: reduce \texttt{WindSigma0} and/or \texttt{WindSpeedFactor} -- more predictable, monotonic response than AGN.
\item \pbbold{Then AGN, cautiously}: reduce \texttt{BlackHoleFeedbackFactor}/\texttt{BlackHoleAccretionFactor}, or raise \texttt{SeedBlackHoleMass}. Given the self-regulating accretion loop, run a local bracketing scan (2-3 values around current setting) before trusting the sign of the response -- don't assume a linear gradient from a single perturbation.
\end{itemize}

\par\medskip\noindent\pbbold{If small-scale power is excessive relative to target (too much power at high k)}\par\smallskip
\begin{itemize}[leftmargin=*,itemsep=1pt,topsep=2pt,parsep=0pt]
\item \pbbold{Check for overcooling first}: if gas is over-condensing under this setup's default cooling settings, that's a fixed-parameter effect you can't directly counter -- but confirm it's not a feedback-strength issue before concluding that.
\item \pbbold{Increase wind strength} (\texttt{WindSigma0}, \texttt{WindSpeedFactor}) as the more predictable lever.
\item \pbbold{AGN strengthening} (raise \texttt{BlackHoleFeedbackFactor}/\texttt{BlackHoleAccretionFactor}, lower \texttt{SeedBlackHoleMass}) plausible but same caveat applies -- verify with a bracketing scan rather than a single trial.
\end{itemize}

\par\medskip\noindent\pbbold{If suppression trough is present but at the wrong k (too broad/narrow, or centered at wrong scale)}\par\smallskip
\begin{itemize}[leftmargin=*,itemsep=1pt,topsep=2pt,parsep=0pt]
\item \pbbold{Adjust \texttt{WindFreeTravelLength}/\texttt{WindFreeTravelDensFac}} -- controls how far ejected gas travels before re-coupling, setting the k-range of wind-driven suppression; increasing \texttt{WindFreeTravelLength} pushes suppression to larger scales (lower k). More reliable scale-control knob.
\item \pbbold{\texttt{SeedBlackHoleMass}} shifts the halo-mass range where AGN feedback activates, setting the k-location of any AGN-driven trough -- treat as a secondary/exploratory knob given the self-regulation caveat, not a precise dial.
\end{itemize}

\par\medskip\noindent\pbbold{If BAO wiggle amplitude/damping is off}\par\smallskip
\begin{itemize}[leftmargin=*,itemsep=1pt,topsep=2pt,parsep=0pt]
\item \pbbold{Adjust \texttt{OmegaBaryon}} (with \texttt{Omega0} fixed, trading off against CDM fraction) -- directly controls baryon-drag suppression and wiggle contrast; increasing OmegaBaryon deepens wiggle damping and increases suppression at k >\textasciitilde{} 0.1 h/Mpc.
\end{itemize}

\par\medskip\noindent\pbbold{If turnover scale (matter-radiation equality feature) is shifted}\par\smallskip
\begin{itemize}[leftmargin=*,itemsep=1pt,topsep=2pt,parsep=0pt]
\item \pbbold{Adjust \texttt{HubbleParam}} -- shifts turnover in h/Mpc units by dk/k \textasciitilde{} dh/h.
\item Secondarily, \texttt{Omega0} also shifts equality redshift and turnover position, but couples more strongly to overall amplitude -- prefer \texttt{HubbleParam} if amplitude away from turnover already matches target.
\end{itemize}

\par\medskip\noindent\pbbold{If results look numerically unconverged (noisy, discontinuous vs. what you'd expect from a smooth parameter change, or high-k power collapses/artificially truncates)}\par\smallskip
\begin{itemize}[leftmargin=*,itemsep=1pt,topsep=2pt,parsep=0pt]
\item Most likely the fixed \texttt{Ngrid}/\texttt{BoxSize} resolution limit (see Non-physical/numerical parameters above), not something you can fix by changing a tunable parameter. Note it and move on rather than chasing it with feedback parameters.
\end{itemize}

\par\medskip\noindent\pbbold{If fit looks good on average but has large scatter/noise at low k specifically}\par\smallskip
\begin{itemize}[leftmargin=*,itemsep=1pt,topsep=2pt,parsep=0pt]
\item Likely \pbbold{sample variance from this trial's \texttt{Seed}} (every trial draws its own -- see TASK.md), not a parameter mis-specification. Don't over-correct a single trial's low-k wiggle; check whether the trend holds up over a couple of trials before concluding your parameters moved things the wrong way.
\end{itemize}

\par\medskip\noindent\pbbold{Suggested optimization priority order (highest confidence / most reliably monotonic first)}\par\smallskip
\begin{enumerate}[leftmargin=*,itemsep=1pt,topsep=2pt,parsep=0pt]
\item \texttt{Sigma8} (overall amplitude) -- high confidence, monotonic
\item \texttt{PrimordialIndex} (large-vs-small-k tilt) -- high confidence, monotonic
\item \texttt{Omega0} / \texttt{OmegaBaryon} (shape near turnover/BAO) -- moderate confidence, use jointly with amplitude fixed
\item \texttt{HubbleParam} (turnover position, fine correction) -- moderate confidence
\item \texttt{WindSigma0} / \texttt{WindSpeedFactor} (small-scale suppression strength) -- moderate confidence, reasonably monotonic
\item \texttt{WindFreeTravelLength} / \texttt{WindFreeTravelDensFac} (small-scale suppression scale/k-location) -- moderate confidence
\item \texttt{BlackHoleFeedbackFactor} / \texttt{BlackHoleAccretionFactor} / \texttt{SeedBlackHoleMass} (small-scale suppression, fine-tuning) -- \pbbold{low confidence without a local sensitivity scan}; treat as exploratory, verify direction empirically before trusting gradient
\end{enumerate}

\pbbold{Note for the optimizer:} for step 7, budget extra evaluations for a small bracketing scan around the current AGN parameter values before computing a gradient-based update, rather than treating a single forward/backward perturbation as reliable. Resolution (\texttt{Ngrid}/\texttt{BoxSize}) is deliberately absent from this list -- it's fixed in this harness, not a tuning target.

\end{promptbox}
\end{document}